\documentclass[conference]{IEEEtran}
\IEEEoverridecommandlockouts
\usepackage{cite}
\usepackage{float}
\usepackage{amsmath,amssymb,amsfonts}
\usepackage{algorithmic}
\usepackage{graphicx}
\usepackage{textcomp}
\usepackage{xcolor}
\usepackage{todonotes}
\usepackage[hyphens]{url}
\usepackage{multirow}
\usepackage[inline]{enumitem}
\usepackage{flushend}
\def\BibTeX{{\rm B\kern-.05em{\sc i\kern-.025em b}\kern-.08em
    T\kern-.1667em\lower.7ex\hbox{E}\kern-.125emX}}
    
\IEEEpubid{\makebox[\columnwidth]{978-1-7281-6680-3/20/\$31.00~\copyright~2020 IEEE \hfill} \hspace{\columnsep}\makebox[\columnwidth]{ }}
\begin{document}

\title{Ransomware as a Service using Smart Contracts and IPFS\\}

\author{\IEEEauthorblockN{Christos Karapapas, Iakovos Pittaras, Nikos Fotiou, George C. Polyzos}
\IEEEauthorblockA{Mobile Multimedia Laboratory\\
Department of Informatics, School of Information Sciences and Technology\\
Athens University of Economics and Business, Greece\\
\{karapapas,pittaras,fotiou,polyzos\}@aueb.gr}
}

\maketitle
\IEEEpubidadjcol
\begin{abstract}
Decentralized systems, such as distributed ledgers and the InterPlanetary File System (IPFS), are designed to offer more open and robust services. However, they also create opportunities for illegal activities. We demonstrate how these 
technologies can be used to launch a ransomware as a service campaign. We show that criminals can transact with affiliates and victims without having to reveal their identity. Furthermore, by exploiting the robustness and resilience to churn of IPFS,
as well as the decentralized 
computing capabilities of Ethereum, criminals can remain offline during most procedures, with many privacy guarantees.
\end{abstract}

\begin{IEEEkeywords}
Blockchain, Distributed Ledger Technologies, Ethereum, InterPlanetary File System
\end{IEEEkeywords}

\section{Introduction}
Ransomware has been a serious problem for Internet users and recently mostly for businesses. Ransomware is malicious software that blocks users from accessing their data (usually by encrypting them) and demands a ransom
to restore access. A recent report from Cybersecurity Ventures predicts that there will be a ransomware attack every 11 seconds by 2021.\footnote{https://cybersecurityventures.com/global-ransomware-damage-costs-predicted-to-reach-20-billion-usd-by-2021/}

Ransomware authors usually take advantage of technologies designed to provide a more open and free Internet. Therefore, it comes as no surprise that they were one of the earliest adopters of \emph{blockchain} technology. Initially, they used cryptocurrencies
to collect the ransom from their victims with high anonymity. However, recent findings show that malware writers leverage blockchain technology as a storage service, or even as a communication channel between them and the ransomware~\cite{blockware}. Nevertheless, 
distributing files over a blockchain can be cumbersome and some times expensive as blockchains are not ideal for storing arbitrary data. 

The InterPlanetary File System (IPFS) is a more attractive alternative for distributing ransomware files. IPFS is a peer-to-peer network that acts as a distributed file system. Being anonymous and robust, IPFS is an attractive platform for malicious activities. Indeed,
in May 2019, Anomali Labs reported the first malware in the wild that used IPFS to hide its command-and-control~(C2) communication~\cite{anomali}. Files stored in IPFS can be easily accessed using HTTPS. Furthermore, files are robustly cached by IPFS nodes; therefore,
they can be retrieved even if the original storage node is offline. 

Currently, cybercrime is a global underground economy with various monetization venues. It is estimated that only in 2018, the revenue of cybercrime was more than \$1.5 billion\cite{mcguire2018into}. Practically, we are facing a huge threat 
that is gradually becoming more organised. Malicious campaigns can be considered more in the scope of enterprises, than of individuals or groups. In fact, as noted multiple times in underground forums, many modern campaigns and attacks are a result 
of a ``service.'' For instance, we have already witnessed the evolution of creating botnets to perform denial of service attacks, to providing the ``botnet-as-a-service,'' i.e., renting compromised hosts to others to perform their attacks. More generally, the case of Malware-as-a-Service is a well-established model, however, as we discuss below, it is still in its infancy. Practically, malware authors ``advertise'' in forums, mostly in the dark Web, or channels of secure messengers,
their skills to develop malware with specific capabilities for a price and usually are paid in a cryptocurrency. 

We argue that while the model works, we expect it to shift to more enterprise status. The shift is expected since it is known that Law Enforcement Agencies (LEAs) and other organizations try to monitor or even backdoor forums and, as a result, reveal the identity of malware
authors. We believe that the key future venue would be further exploitation of decentralized models. Hence, a model where the various functionalities are split, operate individually, and are orchestrated through
a blockchain is a viable alternative that we will face more in the near future. The research question is therefore to predict this trend and identify weaknesses and gaps to be used to counter such threats.

In this paper, we realize the concept of \textit{Ransomware as a Service}~(RaaS). RaaS is, in essence, a ``ransomware affiliation program'': affiliates spread ransomware to potential victims who, upon infection, pay an amount to the ransomware author in exchange
for a (file) decryption key; the affiliate who performed the infection receives a percentage of the ransom and the remainder is received by the RaaS system owner. 

In the considered system, we fully utilize blockchain and IPFS technologies. We take advantage of the Ethereum blockchain by using smart contracts as a registration and ransom payment service for robust, low cost, fair and anonymous transactions.
We also leverage IPFS to reliably host Web pages used for interacting with Ethereum smart contracts, as well as the malicious executable files. This system achieves the following:
\begin{enumerate*}
    \item[(1)] The amount of time a ransomware author needs to be online is minimal.
    \item[(2)] The ransomware author does not have to have a stable network address.
    \item[(3)] The identities of the ransomware authors and the affiliates are hidden. 
    \item[(4)] Affiliates do not have to pay money upfront; instead, malware authors receive a commission from the ransom. 
    \item[(5)] Once up and running, it is too hard to take offline the RaaS system, as well as the registration and payment systems.
\end{enumerate*}

While our research investigates the potential issues of an extension of the current ransomware schemes, we have not detailed important technical aspects that could allow a perpetrator to release the proposed scheme in the wild.
This is aligned with the ethical boundaries set for offensive security research that enable the assessment of a threat, yet prevents its use in the wild before prompt awareness and deployment of precaution measures.

The remainder of this paper is structured as follows: In Section II, we present related work in the area. In Section III, we introduce our blockchain-based architecture and in Section IV, we present its implementation. We evaluate our system in Section V.
Finally, in Section VI, we discuss various aspects of the system and potential mitigation directions and in Section VII we provide conclusions and an outlook.
 
\section{Related work}

Al-rimy et al. \cite{al2018ransomware} have done a very in-depth review of the research on typical ransomware. They also propose a taxonomy from three perspectives: 
severity, platform, and target. However, the authors do not take into consideration the case of ransomware as a service.

Lately, there has been growing interest in blockchain technologies and the different ways they can be exploited for threats and crimes. Moubarak et al. \cite{krymal} studied the blockchain potential in K-ary malware. A K-ary malware, instead of holding virus 
instructions in a single file, consists of K parts whose union results in the malware. Each one of these parts individually is a seemingly benign executable. In their paper, they created a 4-ary malware and used blockchain technology with Proof of Existence~(PoE). 
PoE is a way for someone to prove that a file existed at a specific time. PoE exploits the \texttt{OP\_RETURN} opcode~\cite{bitcoin} to insert into the blockchain the fileÕs hash value~\cite{crosby2016blockchain}. The same opcode is also used in the work of Ali et 
al.~\cite{Ali2018}. The authors leveraged Bitcoin technology to implement their Proof of Concept (PoC) botnet named ZombieCoin. Contrary to these works, which use the blockchain to store the malware, we consider malware storage on the IPFS. We opted for the use of IPFS for file distribution
since it does not require any specialized software, it has low communication overhead and it does not involve monetary cost.  

In addition, previous research tries to exploit blockchains to facilitate the communication between the ransomware and its author. The goal of these efforts is to prevent network administrators from applying takedown mechanisms and allow malware to continue 
their interactions with its creator. Pletinckx et al.~\cite{coordination} proposed a solution that allows ransomware to coordinate using information stored in the Bitcoin blockchain, a technique which is actively used by the Cerber ransomware \cite{cerber}.
Patsakis and Casino~\cite{Patsakis2019} proposed an IPFS-based approach for the same purpose. Our work does not consider any direct communication between the ransomware and its author; instead, these entities communicate with each other 
indirectly through the deployed smart contracts.

One of the main reasons for the rise of ransomware is the adoption of cryptocurrencies. This enabled malware authors to receive the ransom in a seamless and anonymous way, leaving few if any traces that could lead to their prosecution. The economic model that is used
by ransomware has already been studied in the literature \cite{laszka2017economics,caporusso2018game,cartwright2019pay}, trying to identify for instance whether it is worth paying the ransom or whether the ransomware could request a higher price. In fact, during
the past few years, we are witnessing less malware, but more targeted to organisations than individuals, in order to maximise revenue.

\section{System Design}

\begin{figure*}
    \centering
    \includegraphics[width=.8\textwidth]{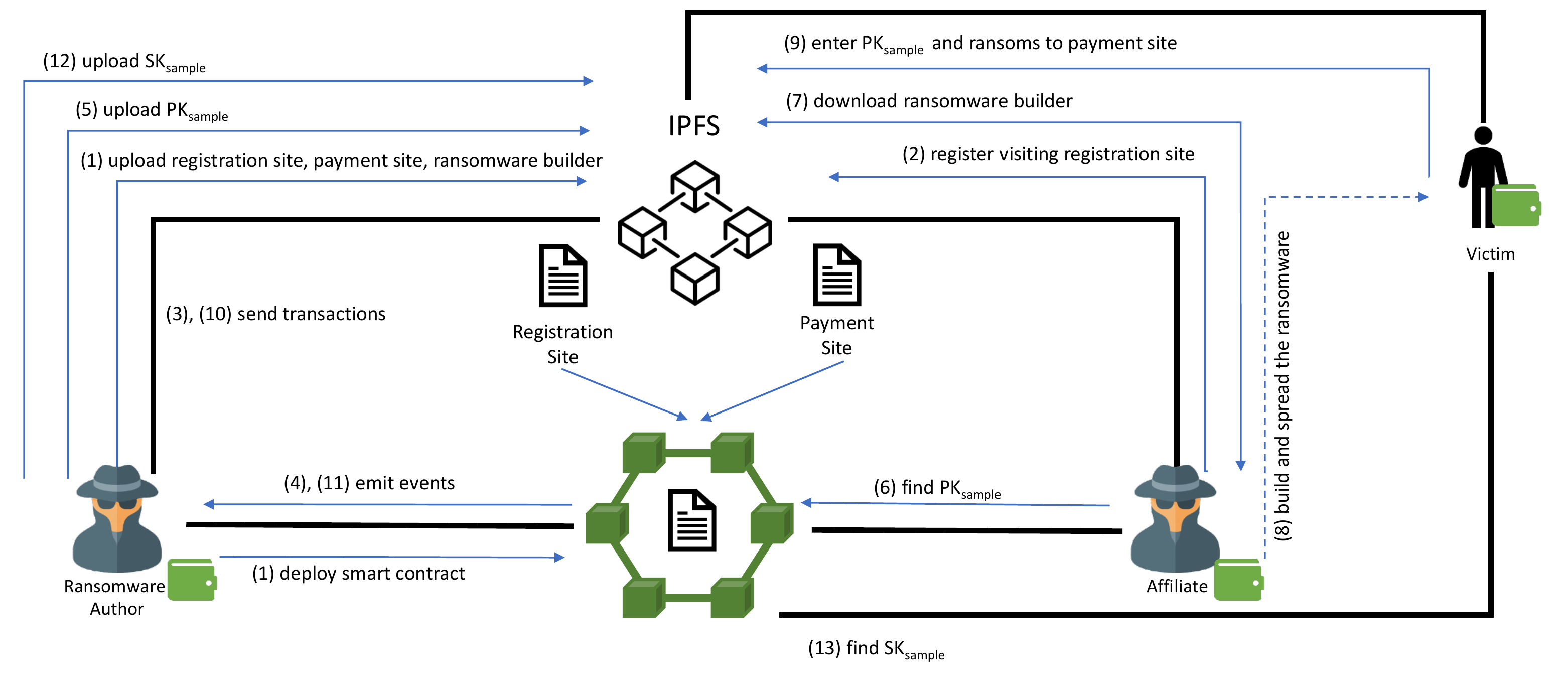}
    \caption {An overview of the considered blockchain-based architecture.}
    \label{fig1}
\end{figure*}

We now present the design of the considered architecture (illustrated in Figure~\ref{fig1}). The system is composed of the following entities: the ransomware \emph{author} who creates the original ransomware, some \emph{affiliates} who buy the ransomware from an author and
try to infect \emph{victims}. From a high-level perspective, these entities interact with each other as follows. Authors store their ransomware in IPFS and make it available using an Ethereum smart contract. Affiliates obtain it and try to infect other users. Infected users
(i.e., the victims) use another Ethereum smart contact to pay the ransom and receive the decryption key. The author and the corresponding affiliate share the ransom. This functionality is implemented using the following steps.   

\subsection{Setup}
During the setup phase, the author creates an Ethereum account and stores it in a wallet. We will refer to the Ethereum address of the author as $A_{address}$.  Then, she creates a \emph{registration} Web page, a \emph{payment} Web page, and a ransomware
\emph{builder}, and uploads them on IPFS. Moreover, she has to create the corresponding smart contracts and deploy them on the Ethereum blockchain. In order for the author to upload the created files in IPFS, she needs an IPFS node running on a computer under her
control, but after the files are received by other IPFS nodes, her local node is no longer required. The registration Web page is used by
affiliates to join the affiliates program, whereas the payment page is used by victims to pay the ransom. Both pages interact with the corresponding smart contracts
and are realized on IPFS for robustness. Finally, the author sets up a  C2 application which interacts with all other components \emph{only 
through the Ethereum smart contracts}. The C2 application is connected to the Ethereum network and it is configured to ``listen'' for specific events. 

\subsection{Affiliate registration}
An affiliate joins the affiliation program through the registration page. During this process, the affiliate registers his Ethereum address, which is stored in the corresponding smart contract, and downloads the builder. For each user he wishes to infect, 
the affiliate uses the builder to create a ransomware \emph{sample}. Whenever a new sample is created, it uses the registration smart contract to request a \emph{public key} $PK_{sample}$. This request creates an event which is broadcasted in the Ethereum network; 
hence it is received by the  C2 application; the C2 application generates a public-private key pair using as seed the transaction hash, and returns the public key to the smart contract. The smart contract verifies that $PK_{sample}$ was sent by $A_{address}$. The 
smart contract stores all $PK_{sample}$ associated with an affiliate address in a data structure.

\subsection{Victim infection}
An affiliate may distribute the ransomware using various mechanisms, e.g. exploiting a system vulnerability, spear phishing, etc.; however, the means to do it are beyond the scope of this work. When the ransomware is executed by a victim, it creates a 
symmetric encryption key $Key_{temp}$, which is used to encrypt some critical for the victim files. Then, the ransomware encrypts $Key_{temp}$ using $PK_{sample}$ and stores the ciphertext locally. Finally, the ransomware notice is presented to the user 
containing the $PK_{sample}$, the URL of the payment page, and the amount of money, in ether, that the victim must pay to get his files decrypted. 

\subsection{Ransom payment}
In order for the victim to get his files decrypted, he has to visit the payment page and enter $PK_{sample}$, as well as to deposit the predetermined amount. This amount is transferred from his Ethereum wallet to the smart contract's balance, 
which in return transfers a proportion of the ransom to the affiliate and the rest to the ransomware author. This action emits an event, which is received by the  C2 application. The application retrieves the corresponding secret key $SK_{sample}$ and sends 
it to the smart contract. The smart contract verifies that $SK_{sample}$ was sent by $A_{address}$. The ransomware uses $SK_{sample}$ to decrypt $Key_{temp}$, and then it uses $Key_{temp}$ to decrypt the files of the victim.   

\section{Implementation}
We developed a proof of concept implementation of the blockchain-based architecture and we deployed it on the Rinkeby test network. Ethers in the Rinkeby network do not have any real value. Furthermore, the deployed system does not contain a true infection mechanism and is thus not directly
usable, i.e., harmless. 

The smart contracts of the system were developed using 
Solidity. 
These contracts implement seven functions, corresponding to the actions of the system. The first function is used for the affiliate registration; it stores the necessary information in the blockchain, and it emits an event when the process is completed. 
Two other functions are used for setting and getting a $PK_{sample}$. A fourth function implements ransom payment, and it is responsible for triggering the corresponding event. A fifth function is responsible for splitting the ransom between the affiliate and 
the author. Finally, there are two functions for setting and getting $SK_{sample}$.

As we have already mentioned, the ransomware author is expected to create the two sites and upload them to IPFS. The sites are implemented as React 
applications with a simple user interface. 
The sites are interacting with the smart contracts, using the web3 JavaScript library. 

Finally, the implementation of this system includes the C2 application that is executed on behalf of the ransomware author. This is a script developed in Node.js.

\section{Evaluation}
\subsection{Performance and Cost Evaluation}
All actions performed in the system involve the invocation of the smart contract functions discussed in the previous section. Two of them only read the state of the blockchain; thus, 
they have no cost, no significant delay, or serious overhead. The deployment of the smart contract in the blockchain network, as well as the cost (measured in gas) for invoking the contract's functions, are shown in Table~\ref{table1}.

\begin{table}[th]
\centering
\begin{tabular}{ |l| c | c |}
\hline
\textbf{Actor} & \textbf{Operation} & \textbf{Cost measured in gas} \\ \hline
\multirow{4}{*} {Ransomware Author}&  Deployment     &     505822  \\
\cline{2-3}
&$PK_{sample}$ Upload &        29881                      \\
\cline{2-3}
&$SK_{sample}$ Upload &     22144                         \\ 
\cline{2-3}
&Ransom Split & 37515                                    \\
\hline
Affiliate & Affiliate Registration   &          22796                   \\
\hline
Victim & Ransom Payment   &       28326                       \\
\hline
\end{tabular}
\vspace{2mm}
\caption{Cost of the construction building blocks}
\label{table1}
\end{table}

Three of those five operations are initiated by the ransomware author, one by the affiliate, and one by the victim. Therefore the malware author is billed with a total cost of $TC =\ 505822 +\ 29881\cdot\rho + 22144\cdot\delta + 37515 \cdot \mu$ gas units, 
where $\rho$ is the number of registrations, $\delta$ is the number of $SK_{sample}$ keys generates, and $\mu$ is the number of payments received. At the time of writing, the equivalent of 1 Ether in fiat currency is \$175,59
 \footnote{\url{https://coinmarketcap.com/currencies/ethereum/}} and the gas price is set to 1Gwei. Assuming that $\rho = \delta = \mu = 100$, the total cost for the ransomware author for 100 registrations and payments will be $\approx \$1,67$ which is 
minimal by any standard.

In addition to the gas cost, Ethereum also adds an execution time overhead related to the time a transaction needs to be mined. This time depends on the gas price, which is the amount of Ether that a user pays per unit of gas. The higher the gas price, the 
faster the transaction will be mined. On average, an operation in Ethereum is executed in $\approx 13$ seconds.

\subsection{Security and Privacy Properties}
The system has some intriguing security and privacy properties. The Ethereum blockchain offers a high degree of anonymity, as it is hard to track an Ethereum address back to its real-world owner. Therefore, authors and affiliates cannot be easily detected. 
Similarly, once the ransomware and the Web pages are stored in IPFS, the author does not have to participate in the IPFS p2p network. 

An author does not have to be constantly online; neither has he to use the same device or network location. An author's C2 application is triggered by Ethereum events; however, these events are broadcast on the whole Ethereum network. The smart contract controls 
who can write $PK/SK_{sample}$. This access control is implemented by examining the Ethereum address of the entity that made the corresponding transaction; hence it does not reveal any real-world information. 

\section{Discussion}
Two key properties of our system are that a) it can easily bootstrap, and b) it is hard to take it offline.

The system can be easily adopted, since it does not require any upfront payment by the affiliate, apart from the cost for interacting with the blockchain. Instead, affiliates share the ransom with the ransomware author. Ransom sharing is done automatically by a 
smart contract. 

IPFS and Ethereum are two robust systems where information is permanently stored. Therefore, once the ransomware and the Web pages are in the IPFS, and the smart contracts are deployed in Ethereum, they cannot be removed. IPFS provides a ``blacklisting'' functionality, 
which is optional. Similarly, smart contracts can only be removed using chain ``forking'', which is an extremely unlikely process. On the other hand, smart contracts cannot be modified; hence they must be carefully designed so as to be bug-free.  
The major drawback, for criminals, of the proposed scheme is the fact that all the transactions are visible from any participating node of the blockchain. Therefore, a Law Enforcement Agency can identify, for instance, when a new affiliate has been recruited, a 
payment has been made, etc. Nevertheless, the same applies to many wallets which have already been identified with criminal activity. This is the reason for the rise of crypto laundry services, which can efficiently anonymize such revenues \cite{BalthasarH17}. 

It is important to mention that field values in smart contracts, even though they are not declared as public, are still accessible. A smart contract's state is visible to anyone using a blockchain explorer~\cite{atzei}. It is also important to note that even if 
the smart contract stores only information about the last registration, someone can find information about the previous registrations on the blockchain. But note that every piece of information sent through the smart contract can be encrypted 
with the recipient's public key so that only he can decrypt it using his private key, locally.

\section{Conclusion}
In this paper, we presented a system, introducing the idea of RaaS, leveraging Ethereum smart contracts and IPFS. This approach leads to enhanced privacy, it automates payments, it has low cost and low overhead, and it is hard to get it offline.

The system can be extended in many ways. For example, it can leverage zero-knowledge proofs and anonymous payments to provide even more effectiveness, privacy and security for the authors and the affiliates. Additionally, the payment of smart contracts can be extended
to be ``more fair,'' by examining the validity of $SK_{sample}$ before making any payments. Finally, our system can be extended with blockchain-based ransomware coordination. 

The goal of this investigation is not to encourage and help ransomware authors. It is to motivate researchers in this area to find solutions for this newly added threat, introduced by 
blockchains and IPFS. These technologies are here to stay; thus, the study of their possible exploitation becomes a necessity in order to find novel countermeasures. 

\section*{Acknowledgment}
We thank Dr. Constantinos Patsakis
(Department of Informatics, University of Piraeus)
for his helpful comments. 

\bibliographystyle{IEEEtran}
\bibliography{IEEEabrv,submitted}
\vspace{12pt}

\end{document}